\documentclass[twocolumn,aps,prl,showpacs,superscriptaddress,floatfix]{revtex4}
\usepackage{dcolumn}
\usepackage{graphicx}
\usepackage{epsf}
\usepackage{amsmath}

\newcommand{\landauO}{{\mathcal O}}
\newcommand{\addrDresden}{Institut f\"{u}r Theoretische Physik, 
Technische Universit\"{a}t Dresden, 01062 Dresden, Germany}
\newcommand{\addrParis}{Laboratoire Kastler Brossel, \'Ecole Normale Sup\'erieure et
Universit\'{e} Pierre et Marie Curie, Case 74\\
4, pl.\ Jussieu, 75005 Paris, France}
\newcommand{\addrGaithersburg}{National Institute of Standards and Technology, 
Mail Stop 8401, Gaithersburg, MD 20899-8401, USA}

\begin{document}


\title{Asymptotic properties of self-energy coefficients}

\author{Ulrich D.~Jentschura}
\affiliation{\addrDresden}\affiliation{\addrParis}

\author{Eric-Olivier Le~Bigot}
\affiliation{\addrParis}\affiliation{\addrGaithersburg}

\author{Peter J.~Mohr}
\affiliation{\addrGaithersburg}

\author{Paul Indelicato}
\affiliation{\addrParis}

\author{Gerhard Soff}
\affiliation{\addrDresden}

\begin{abstract}
We investigate the asymptotic properties of higher-order binding
corrections to the one-loop self-energy of excited states
in atomic hydrogen. We evaluate the historically problematic
$A_{60}$ coefficient for all P states with principal
quantum numbers $n \leq 7$ and D states
with $n \leq 8$ and find that a satisfactory
representation of the $n$-dependence of the coefficients
requires a three-parameter fit. For the high-energy
contribution to $A_{60}$, we find exact formulas.
The results obtained are relevant for the interpretation
of high-precision laser spectrocopic measurements.
\end{abstract}


\maketitle

Bound-state quantum electrodynamics (QED) occupies a unique position 
in theoretical physics in that it combines all conceptual 
intricacies of modern quantum field theories, augmented
by the peculiarities of bound states, with the experimental
possibilities of ultra-high resolution laser spectroscopy. Calculations
in this area have a long history, and the current status of theoretical
predictions is the result of continuous effort. 
The purpose of this Letter is twofold: first, to present 
improved evaluations of higher-order binding corrections
to the bound-state self-energy for a large number of 
atomic states, including highly excited states with 
a principal quantum number as high as $n = 8$, 
and second, to analyze the asymptotic 
dependence of the analytic results on the bound-state quantum
numbers.
Highly-excited states (e.g, with $n=4$ to $12$) are of particular
importance for high-precision spectroscopy experiments in
hydrogen (for a summary, see for instance~\cite[p.~371]{MoTa2000}).

%
\begin{figure}[htb]%
\begin{center}\includegraphics[width=0.7\linewidth]{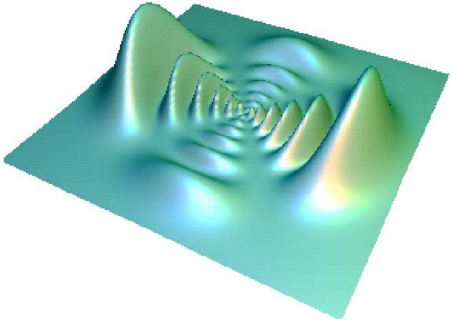}\end{center}%
\caption{\label{fig1}%
Plot of the radial 
probability density $r^2 |\psi(r,\theta,\phi)|^2$ of the 
nonrelativistic 8D wave function (angular momentum
projection $m=0$) in the plane of constant 
azimuth $\phi = 0$. The calculation of the 
relativistic Bethe logarithm $A_{60}$ starts
from this nonrelativistic wave function, 
with relativistic corrections being taken into 
account via generalized Foldy-Wouthuysen 
transformations~\cite{JePa1996,JeSoMo1997}.}%
\end{figure}

In the analytic calculations, we focus on a specific 
higher-order binding correction, known as the 
$A_{60}$ coefficient or ``relativistic Bethe logarithm.''
We write the (real part of the) one-loop self-energy shift
of an electron in the field of a nucleus of charge number~$Z$ as
\begin{equation}
\label{ESEasF}
\Delta E_{\rm SE} = \frac{\alpha}{\pi} \, \frac{(Z \alpha)^4}{n^3} \,
F(nl_j,Z\alpha) \, m \, c^2\,,
\end{equation}
where $F(nl_j,Z\alpha)$ is a dimensionless quantity.
In this Letter, we use natural
units with $\hbar = c = m = 1$ and $e^2 = 4\pi\alpha$
($m$ is the electron mass).
The notation $nl_j$ is inspired by the usual spectroscopic
nomenclature: $n$ is the level number, $j$ is the total angular momentum and $l$ is the orbital angular momentum.

The semi-analytic expansion of $F(nl_j,Z\alpha)$
about $Z\alpha = 0$ for a general atomic state with quantum numbers $n$,
$l \geq 1$ and $j$ gives rise to the expression,
\begin{eqnarray}
\label{defFLO}
F(nl_j,Z\alpha) & = & A_{40}(nl_j) + (Z \alpha)^2 \,
\left[A_{61}(nl_j) \, \ln(Z \alpha)^{-2} \right.
\nonumber\\[2ex]
& & \left. + G_{\rm SE}(nl_j,Z\alpha) \right] 
\qquad (l \geq 0)\,,
\end{eqnarray}
where $G_{\rm SE}(nl_j,Z\alpha) \to {\rm constant}$ as
$Z\alpha \to 0$.
The limit as $Z\alpha \to 0$ of $G_{\rm SE}(nl_j,Z\alpha)$ is referred
to as the $A_{60}$ coefficient, i.e.
\begin{equation}\label{eq:A60}
A_{60}(nl_j) = \lim_{Z\alpha \to 0} G_{\rm SE}(nl_j, Z\alpha)\,.
\end{equation}
It is this coefficient which has proven to be by far the 
most difficult to evaluate~\cite{ErYe1965a,ErYe1965b,
Er1971,Sa1981,Pa1993,JeMoSo1999}. Furthermore, the complexity of the calculation
increases sharply with increasing principal quantum number $n$,
both due to the more involved structure of the bound-state
wave function (see also Fig.~\ref{fig1}), and due to the necessity
of subtracting bound-state poles that lie infinitesimally close to the
photon integration contour. 
The atomic states with the highest $n$ for which analytic results are available 
today are the 4P states~\cite{JeSoMo1997}.
In this Letter, we present analytic data for the $A_{60}$ coefficient of P states with 
$n \leq 7$ and all D states with $n \leq 8$.
For a given $n$, the calculation is more involved
for $n{\rm P}$ than for $n{\rm D}$,
because there is one more term in the nonrelativistic 
radial $n{\rm P}$ wave function than in the corresponding $n{\rm D}$ wave function (when they are expressed as a function of the electron--nucleus distance).
Essentially, the number of terms 
in the radial wave function determines the complexity of the calculation.

One of the most demanding specific calculations in the evaluation
of $A_{60}$ is necessitated by a Bethe-logarithm type contribution
given by the relativistic wave-function correction $F_{\delta \phi}$;
this contribution is defined in Eqs.~(43) and (53) of~\cite{JePa1996}.
For 7P and 8D states, we use up to 200,000 terms in intermediate
steps in the evaluation of this correction.
Because $A_{60}$ involves relativistic corrections to the
coefficient $A_{40}$, which in turn is given mainly
by the Bethe logarithm, it is natural to refer
to the entire $A_{60}$ coefficient as a ``relativistic Bethe logarithm.''

%
%
\begin{table*}[htb]
\caption{\label{table1} $A_{60}$ coefficients for ${\rm P}_{1/2}$,
${\rm P}_{3/2}$, ${\rm D}_{3/2}$ and ${\rm D}_{5/2}$ states 
($n=2,\dots,7$ for P states, and $n=3,\dots,8$ for D). 
The quantity ${\cal L}$ is implicitly
defined in Eq.~(\ref{LowGen}) and represents the low-energy contribution
to $A_{60}$.}
\begin{ruledtabular}
\newcolumntype{e}{D{.}{.}{15}}
\begin{tabular}{leeee}
$n$ & 
\multicolumn{1}{c}{$A_{60}(n{\rm P}_{1/2})$} &
\multicolumn{1}{c}{${\cal L}(n{\rm P}_{1/2})$} &
\multicolumn{1}{c}{$A_{60}(n{\rm P}_{3/2})$} &
\multicolumn{1}{c}{${\cal L}(n{\rm P}_{3/2})$} \\
\hline
2 \hspace{0.2in} &
  -0.998~904~402(1) & 
  -0.795~649~812(1) &
  -0.503~373~464(1) & 
  -0.584~516~780(1) \\
3 & 
  -1.148~189~956(1)\footnote{We take the opportunity to correct
a computational error for this result as previously reported in
Ref.~\cite[Eq.~(96)]{JeSoMo1997}, where a value of $-1.14768(1)$ was given.} &
  -0.944~288~447(1) &
  -0.597~569~388(1) &
  -0.693~566~427(1) \\
4 & 
  -1.195~688~142(1) &
  -0.997~810~211(1) &
  -0.630~945~796(1) &
  -0.730~579~137(1) \\
5 & 
  -1.216~224~512(1) &
  -1.023~991~781(1) & 
  -0.647~013~509(1) &
  -0.747~615~653(1) \\
6 & 
  -1.226~702~391(1) &
  -1.039~079~399(1) &
  -0.656~154~893(1) &
  -0.756~897~499(1) \\
7 &
  -1.232~715~957(1) &
  -1.048~800~134(1) &
  -0.662~027~568(1) &
  -0.762~622~956(1) \\
\hline
\hline
$n$ & 
\multicolumn{1}{c}{$A_{60}(n{\rm D}_{3/2})$} &
\multicolumn{1}{c}{${\cal L}(n{\rm D}_{3/2})$} &
\multicolumn{1}{c}{$A_{60}(n{\rm D}_{5/2})$} &
\multicolumn{1}{c}{${\cal L}(n{\rm D}_{5/2})$} \\
\hline
3 \hspace{0.2in} &
   0.005~551~575(1) &
   0.021~250~354(1) & 
   0.027~609~989(1) &
   0.019~188~397(1) \\
4 &
   0.005~585~985(1) &
   0.022~882~528(1) &
   0.031~411~862(1) &
   0.020~710~720(1) \\
5 &
   0.006~152~175(1) &
   0.023~759~683(1) & 
   0.033~077~571(1) &
   0.021~511~798(1) \\
6 &   
   0.006~749~745(1) &
   0.024~294~690(1) &
   0.033~908~493(1) &
   0.021~975~925(1) \\
7 &
   0.007~277~403(1) &
   0.024~645~479(1) &
   0.034~355~926(1) &
   0.022~264~036(1) \\
8 &   
   0.007~723~850(1) &
   0.024~886~986(1) &
   0.034~607~492(1) &
   0.022~452~259(1) \\
\end{tabular}
\end{ruledtabular}
\end{table*}

The ``normal Bethe logarithm'' $\ln k_0(nl)$ forms part
of the coefficient $A_{40}$ for which a well-known general formula 
(see, e.g., Ref.~\cite[p.~468]{MoTa2000}) reads
\begin{equation}
\label{A40gen}
A_{40}(nl_j) = -\frac{1}{2\kappa\,(2 l + 1)} - \frac{4}{3}\,\ln k_0(nl)\,,
\end{equation}
where $\kappa = 2\,(l-j)\,(j+1/2)$.
Formulas for $A_{61}$ valid for P and D states read as 
follows (see, e.g., Ref.~\cite[p.~468]{MoTa2000})
\begin{eqnarray}
\label{A61P12}
A_{61}(nP_{1/2}) &=& \frac{1}{45}\, \left(33 - \frac{29}{n^2}\right)\,,
\\[2ex]
\label{A61P32}
A_{61}(nP_{3/2}) &=& \frac{2}{45}\, \left(9 - \frac{7}{n^2}\right)\,,
\\[2ex]
\label{A61L2}
A_{61}(nl_j) &=&
\frac{32 \, \left(3 - \frac{\displaystyle l\,(l+1)}{\displaystyle n^2} \right)}
{3 \prod\limits_{m=-1}^3 (2\,l + m)} \quad
(l \geq 2)\,.
\end{eqnarray}
Note that $A_{61}(nl_j) \to {\rm constant}$ for $n \to \infty$ at
constant $l$ and $j$.
It is the purpose of this Letter to present new results for the $A_{60}$ coefficients.  Details of our calculations will be presented in
a forthcoming article~\cite{lebigot2003c}.
It has been observed previously by Karshenboim~\cite{Ka1997priv} that the 
$n$-dependence of the $A_{60}(n{\rm P})$ coefficients can be fitted
to a satisfactory accuracy by an $(n^2-1)/n^2$-type model,
and a two-parameter fit has been employed for the 
$n$-dependence of the S-state coefficients
$A_{60}(n{\rm S}_{1/2})$~\cite{Ka1997}.
Our data for P states in Tab.~\ref{table1} are roughly
consistent with this $(n^2-1)/n^2$ model.

\begin{figure}[htb]
\begin{center}
%
\parbox{4.2cm}{
\centerline{\mbox{\epsfysize=5.0cm\epsffile{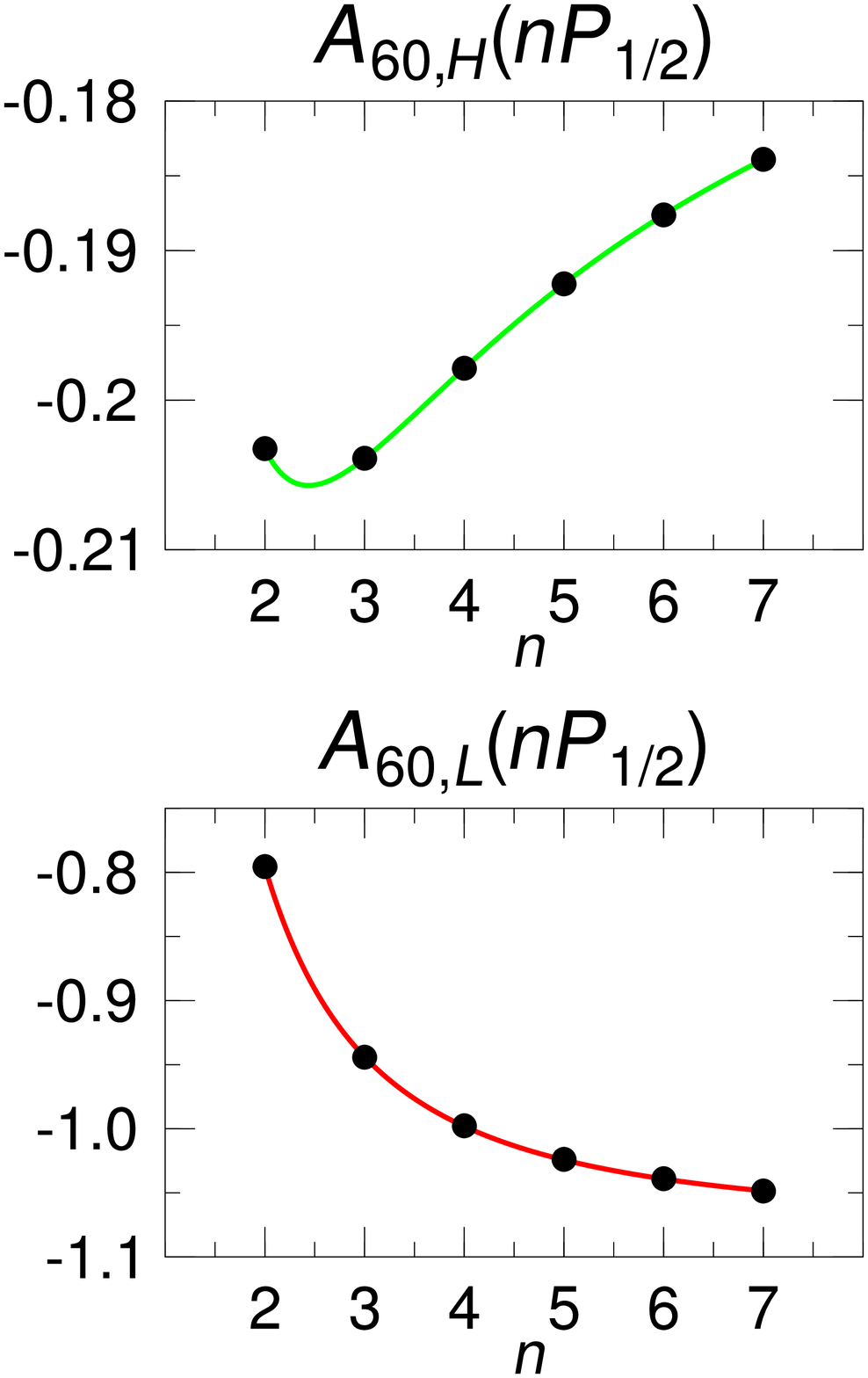}}}
\vspace{0.2cm}%
\centerline{\ \mbox{\epsfysize=2.4cm\epsffile{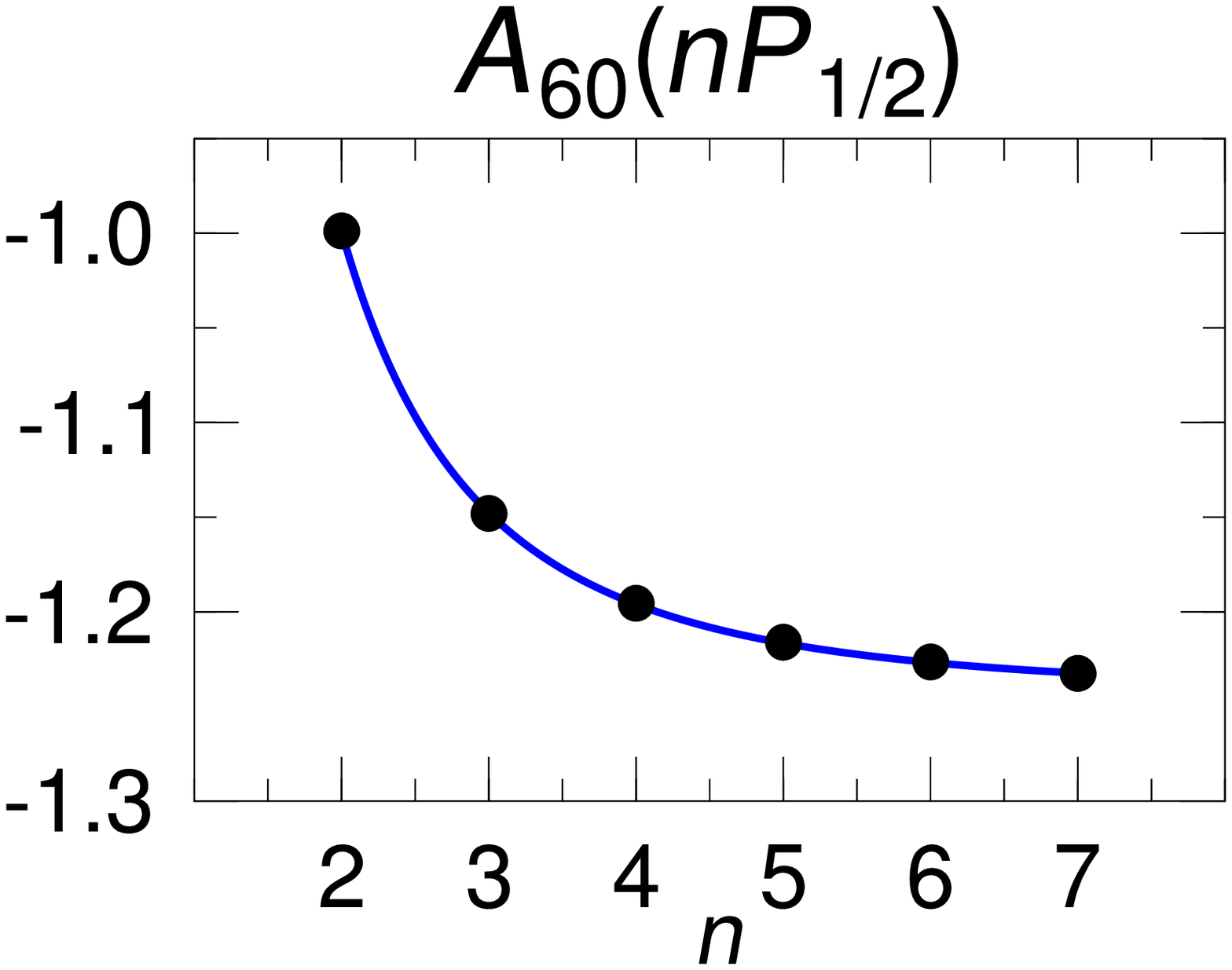}}}
}
\parbox{4.2cm}{
\centerline{\mbox{\epsfysize=5.0cm\epsffile{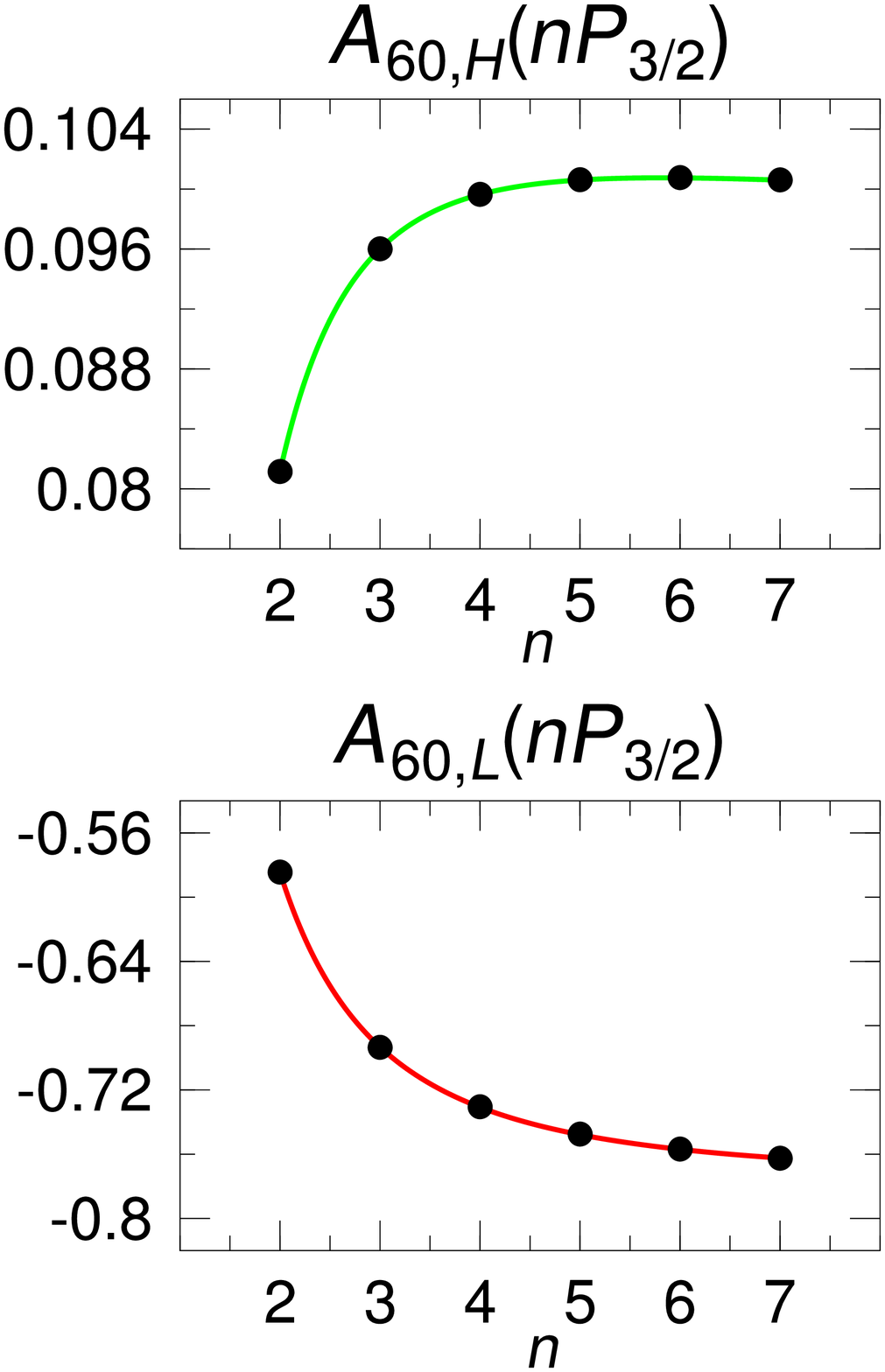}}}
\vspace{0.2cm}%
\centerline{\ \mbox{\epsfysize=2.45cm\epsffile{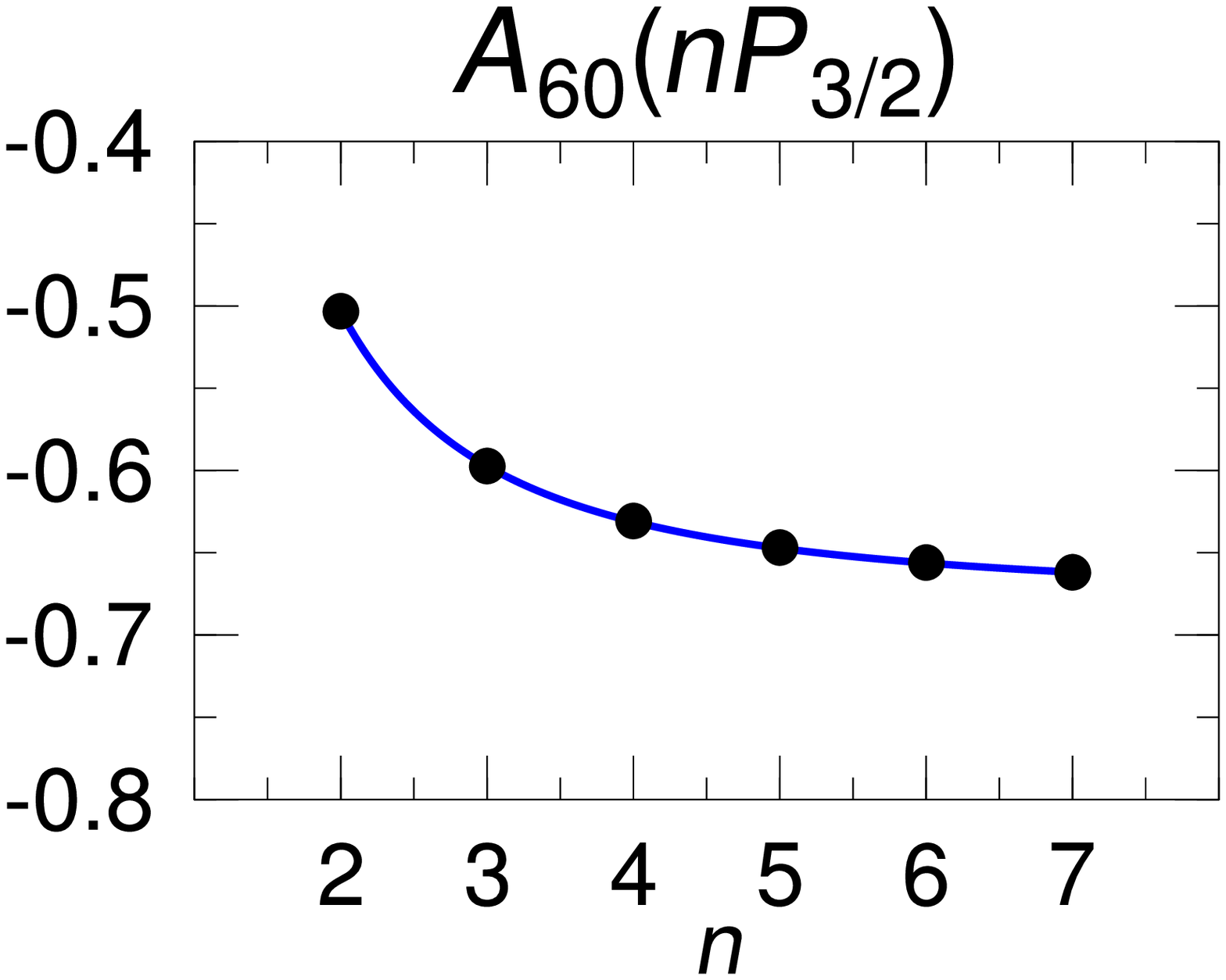}}}
}
\caption{\label{fig2} 
The plots show the dependence on the principal quantum 
number~$n$ of the high- and low-energy part of the~$A_{60}$ 
self-energy coefficient, as well as their sum ($A_{60}$).
The curves for the high-energy contribution $A_{60,H}$
represent the exact results (\ref{A61P12})--(\ref{A61L2}), (\ref{A60H}),
and (\ref{Kgen}), with $n$ being generalized to a continuous
variable (only integer $n$ values have physical significance).
The smooth curves
for the low-energy parts $A_{60,L} = {\cal L}$
result from a three-parameter fit of the data in 
Tab.~\ref{table1} to the function in~(\ref{L1L2L3});
 the fit parameters are
given in Tab.~\ref{table2}.
The curves in the lower row represent 
the total result for
$A_{60} = A_{60,H} + A_{60,L}$.}
%
\end{center}
\end{figure}

\begin{figure}[htb]
\begin{center}
%
\parbox{4.2cm}{%
\hspace*{-0.15cm}%
\centerline{\mbox{\epsfysize=5.0cm\epsffile{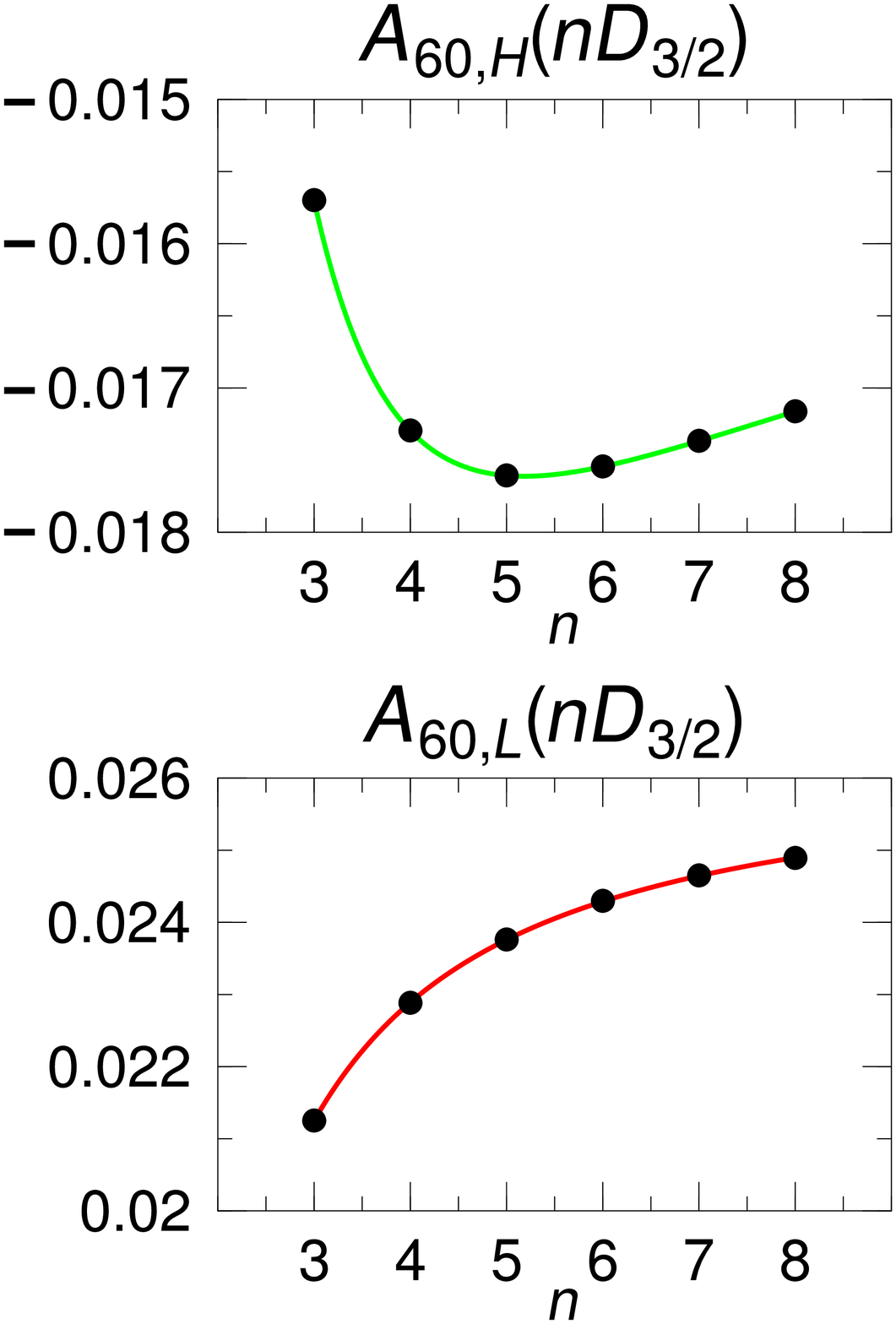}}}%
\vspace{0.2cm}
\centerline{\mbox{\epsfysize=2.4cm\epsffile{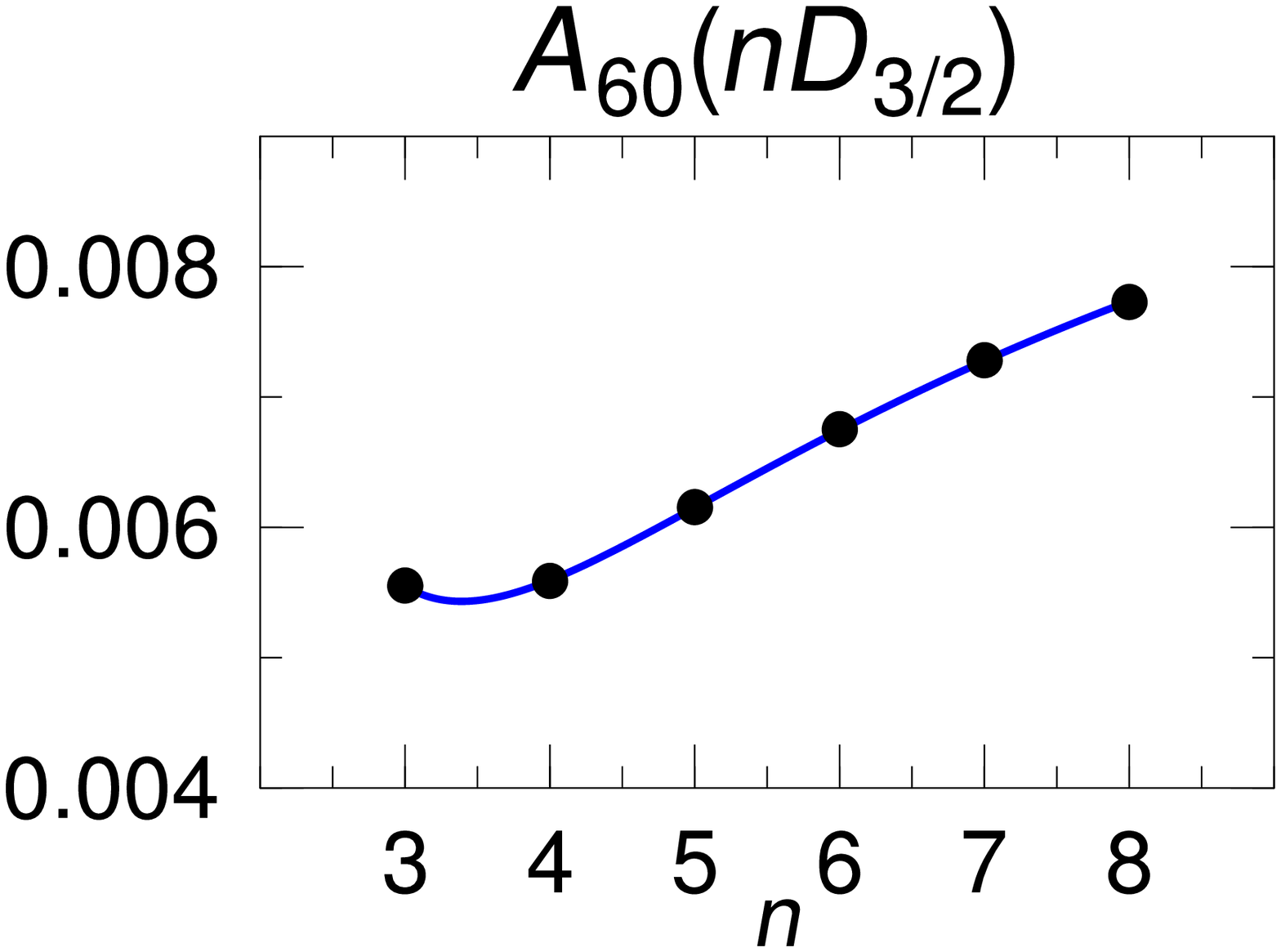}}}%
}
\parbox{4.2cm}{%
\hspace*{0cm}%
\centerline{\mbox{\epsfysize=5.0cm\epsffile{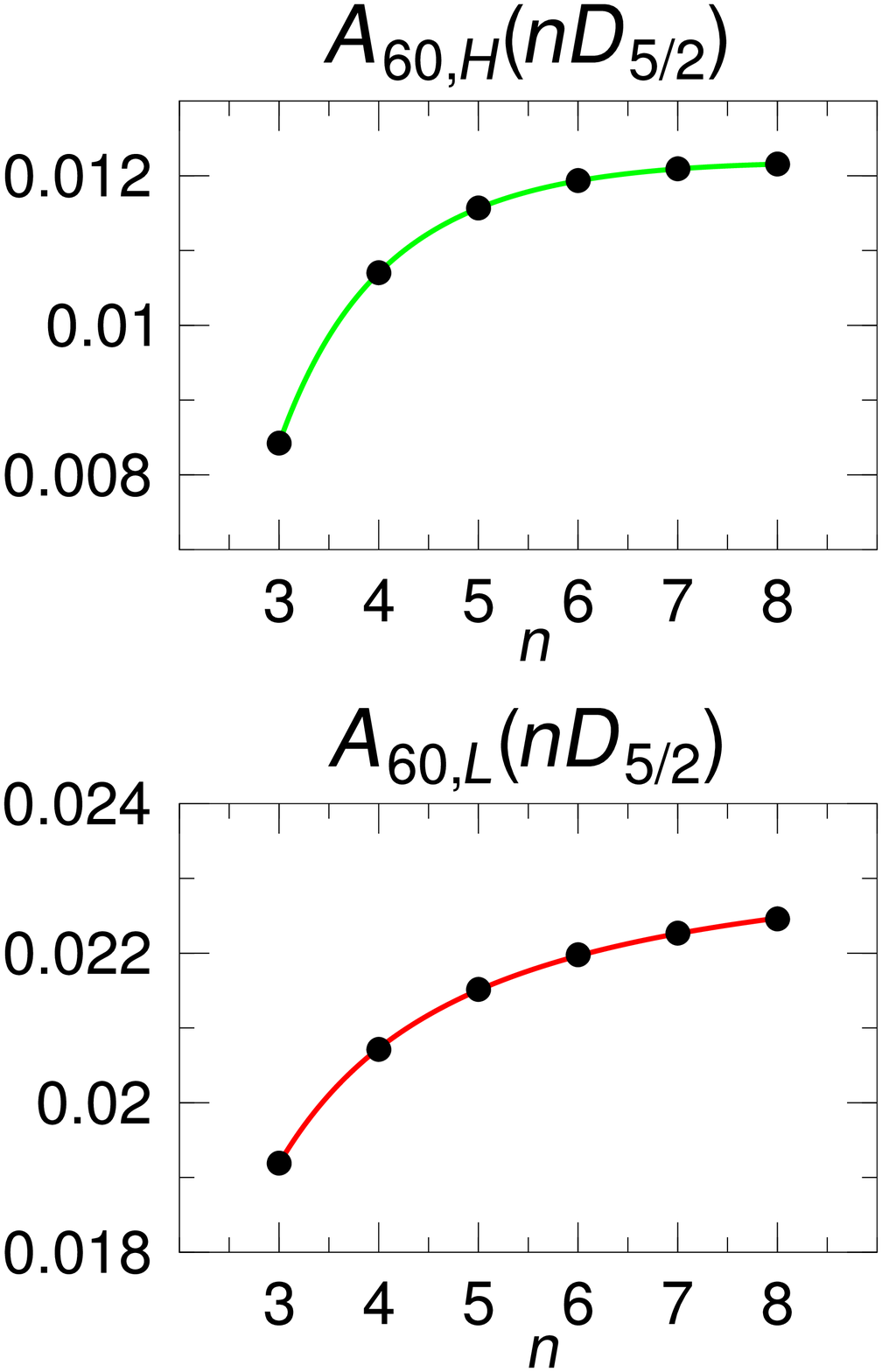}}}%
\vspace{0.2cm}
\centerline{\mbox{\epsfysize=2.4cm\epsffile{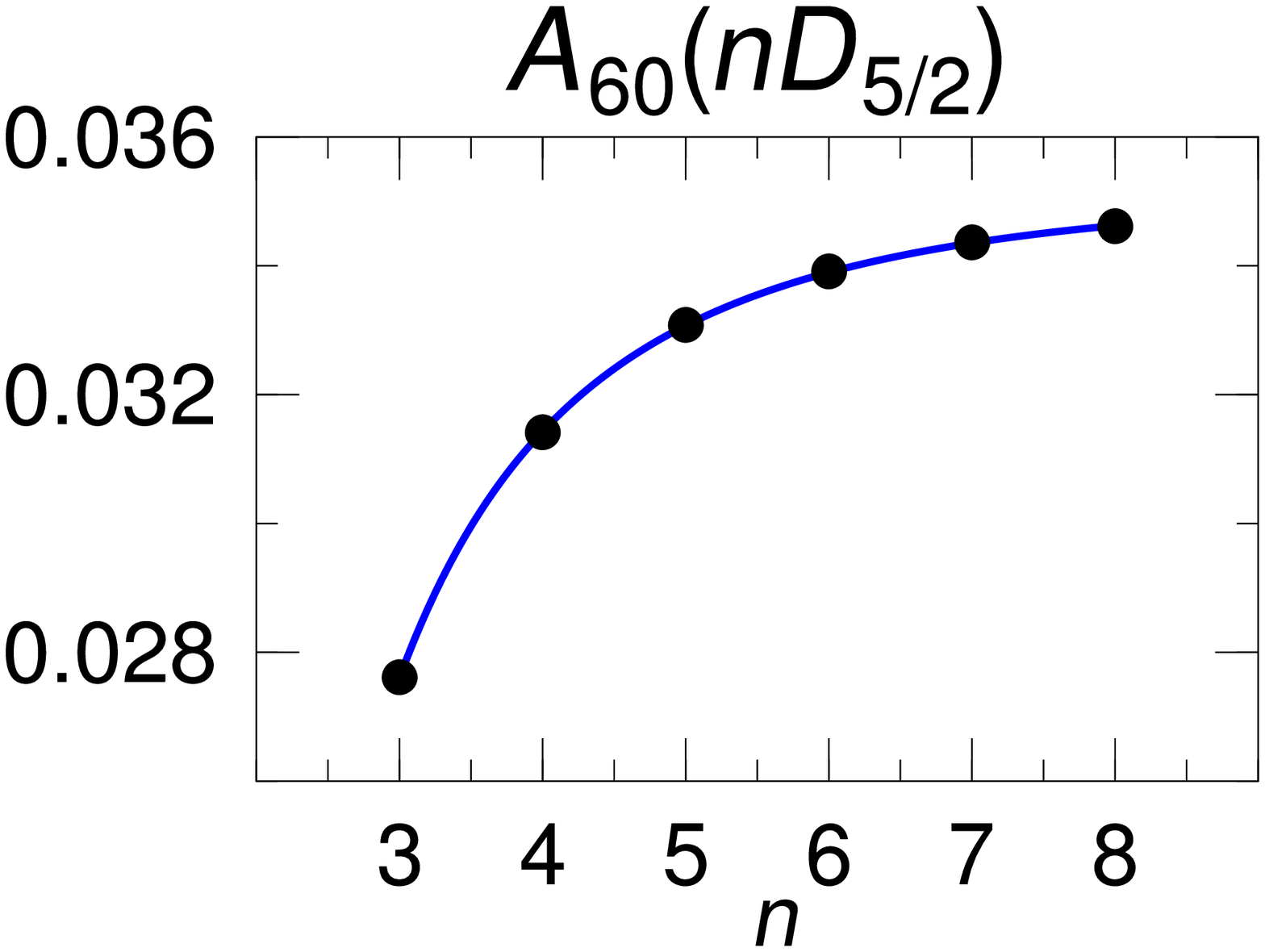}}}
}
\caption{\label{fig3} The analog of Fig.~\ref{fig2}
for ${\rm D}_{3/2}$ and ${\rm D}_{5/2}$ states.
The minimum in $A_{60,H}(n{\rm D}_{3/2})$ near $n=5$ 
is determined by the exact formulas (\ref{A61L2}) and~(\ref{KgenD32}).}
%
\end{center}
\end{figure}

For the atomic states under investigation,
the self-energy contribution due to hard virtual photons
(high-energy part) obtained by the 
$\epsilon$ method~\cite{Pa1993,JePa1996,JeSoMo1997,lebigot2003c}
is
\begin{eqnarray}
\label{HighGen}
\lefteqn{F_H (nl_j, Z\alpha) = - \frac{1}{2\kappa\,(2 l + 1)}}&&
\\\nonumber
& & \quad {}+ (Z\alpha)^2 \,
  \left[{\cal K} - \frac{{\cal C}}{\epsilon} -
    A_{61} \, \ln(2\epsilon) + \landauO(\epsilon) \right]
+ \ldots
\end{eqnarray}
The ellipsis denotes higher-order terms, which are irrelevant 
for the current investigation.
In Eq.~(\ref{HighGen}), $\cal K$ and ${\cal C}$, as well as
$A_{61}$, are state-dependent coefficients.
For concrete evaluations of the high-energy
part concerning specific atomic states, 
see~\cite[Eqs.~(18) and~(19)]{JePa1996}
and~\cite[Eqs.~(55)--(58)]{JeSoMo1997}.
The low-energy part assumes the form
\begin{eqnarray}
\label{LowGen}
\lefteqn{F_L (nl_j, Z\alpha)  =  - \frac{4}{3} \ln k_0(nl)}
\quad\quad&&
\\  \nonumber
& & {}+ (Z\alpha)^2 \, \left[{\cal L} + \frac{{\cal C}}{\epsilon} +
A_{61} \, \ln\left(\frac{\epsilon}{(Z \alpha)^2}\right)
+ \landauO(\epsilon) \right]\,,
\end{eqnarray}
where we omit terms that are irrelevant at relative order $(Z\alpha)^2$
in the evaluation of $F(nl_j,Z\alpha)$.
A detailed explanation of the $\epsilon$ method will be given
in~\cite{lebigot2003c}.
The dependence on ${\cal C}$ cancels when the high- and low-energy parts 
are added. Specifically, we have
\begin{equation}
\label{A60KL}
A_{60} = {\cal K} - A_{61} \, \ln 2 + {\cal L}\,.
\end{equation}
Upon inspection of (\ref{HighGen}) and (\ref{LowGen}),
we identify 
\begin{equation}
\label{A60H}
A_{60,H} = {\cal K} - A_{61} \, \ln 2
\end{equation}
as the high-energy contribution to $A_{60}$,
and $A_{60,L} = {\cal L}$ as
the low-energy contribution (see Figs.~\ref{fig2} and~\ref{fig3}).

%
%
\begin{table}[htb]
\caption{\label{table2} Coefficients $L_1$, $L_2$ and 
$L_3$ that result from a least-squares fit of the 
$n$-dependence of our data for ${\cal L}$ in Tab.~\ref{table1}
(see also Figs.~\ref{fig2} and~\ref{fig3}). The value of $L_1$ from this global fit should approximate the limit $\lim_{n\rightarrow\infty} {\cal L}(nl_j)$ in Eq.~(\ref{L1L2L3}), although it is not necessarily the best estimate.} 
\begin{ruledtabular}
\newcolumntype{f}{D{.}{.}{5}}
\begin{tabular}{lfff}
state &
\multicolumn{1}{c}{$L_1$} &
\multicolumn{1}{c}{$L_2$} &
\multicolumn{1}{c}{$L_3$} \\
\hline
${\rm P}_{1/2}$ \hspace{0.2in} &
  -1.082 &
   0.0966 & 
   0.950 \\
${\rm P}_{3/2}$ &
  -0.775 &
  -0.0232 &
   0.811\\
${\rm D}_{3/2}$ &
   0.0264 &
  -0.00952 &
  -0.0175 \\
${\rm D}_{5/2}$ &
   0.0235 &
  -0.00568 &
  -0.0220\\
\end{tabular}
\end{ruledtabular}
\end{table}

%
%

We obtain the following general formulas for ${\cal K}$:
\begin{subequations}
\label{Kgen}
\begin{eqnarray}
\label{KgenP12}
{\cal K}(n{\rm P}_{1/2}) &=& \frac{637}{1800} -
\frac{1}{4 n} - \frac{767}{5400 \, n^2}\,,\\[2ex]
\label{KgenP32}
{\cal K}(n{\rm P}_{3/2}) &=& \frac{2683}{7200} +
\frac{1}{16 n} - \frac{2147}{5400 \, n^2}\,,\\[2ex]
\label{KgenD32}
{\cal K}(n{\rm D}_{3/2}) &=& -\frac{157}{30240} -
\frac{3}{80 n} + \frac{3007}{37800 \, n^2}\,,\\[2ex]
\label{KgenD52}
{\cal K}(n{\rm D}_{5/2}) &=& \frac{379}{18900} +
\frac{1}{60 n} - \frac{1759}{18900 \, n^2}\,.
\end{eqnarray}
\end{subequations}
All of these formulas are consistent with a 
limit ${\cal K} \to {\rm constant}$ as $n \to \infty$ 
for constant $l$ and $j$.
The $n$-dependence of the nonrelativistic 
${\cal L}(nl_j)$ contributions as listed in Tab.~\ref{table1}
can be approximated very well using a three-parameter
fit inspired by the above structure found for the 
high-energy ${\cal K}$ contributions. We find 
\begin{equation}
\label{L1L2L3}
{\cal L}(nl_j) \approx L_1(l_j) + \frac{L_2(l_j)}{n} + \frac{L_3(l_j)}{n^2}\,,
\end{equation}
where $L_1$, $L_2$, and $L_3$ assume values 
as listed in Tab.~\ref{table2}
for the series of states under investigation.
The $n$-dependence of the low-energy contributions ${\cal L}$ is
smoother than the corresponding curves for the 
high-energy part (see Figs.~\ref{fig2} and~\ref{fig3}).
The excellent agreement of the fits with the numerical
values of $A_{60,L}$, together with our exact results
for the high-energy part as given by
Eqs.~(\ref{A61P12})--(\ref{A61L2}), (\ref{A60H}),
and (\ref{Kgen}), could suggest a constant limit
of $A_{60}(nl_j)$ as $n\to \infty$ for constant
$l$ and $j$.

%
%

For Rydberg states with the highest $l$
possible for given $n$ (i.e., $l = \bar{l} = n-1$),
our results are consistent with
\begin{equation}
\lim_{n \to \infty} A_{60}(n\bar{l}_j) =  0
\text{ for }
\bar{l}=n-1,
\, j=n-1\pm 1/2
\,
,
\end{equation}
which is plausible to suggest as a conjecture. 
The conjecture is indicated by the trend in the numbers
($j = \bar{l} - 1/2$)
\begin{itemize}
\item $A_{60}(3{\rm D}_{3/2}) = 0.005~551~573(1)$,
\item $A_{60}(4{\rm F}_{5/2}) = 0.002~326~988(1)$,
\item $A_{60}(5{\rm G}_{7/2}) = 0.000~814~415(1)$,
\end{itemize}
as well as in the results ($j = \bar{l}+1/2$)
\begin{itemize}
\item $A_{60}(3{\rm D}_{5/2}) = 0.027~609~989(1)$,
\item $A_{60}(4{\rm F}_{7/2}) = 0.007~074~961(1)$,
\item $A_{60}(5{\rm G}_{9/2}) = 0.002~412~929(1)$.
\end{itemize}
The magnitude of $A_{60}(n \bar{l}_j)$ appears to 
decrease faster than $1/n$. 
In general, relativistic corrections acquire at least
one more inverse power of $n$ when $n=l+1$, $j = n-1 \pm 1/2$,
and $n$ large,
than S or P states of the same $n$. This can for example be seen in the 
relativistic correction of order~$(Z\alpha)^4$ to the Schr{\"o}dinger-Coulomb electron energy [Eq.~(2-87) of~\cite{ItZu1980}],
\begin{eqnarray*}
E_{nj} &=& m - \frac{(Z\alpha)^2\,m}{2 n^2} \\[2ex]
& & {} - \frac{(Z\alpha)^4 \, m}{n^3} \,  \left[ \frac{1}{2 j + 1} +
\frac{3}{8 \, n} \right] + {\landauO}[(Z\alpha)^6].\nonumber
\end{eqnarray*}
For $j = n-1 \pm 1/2$, this relativistic term
acquires an additional inverse power of $n$.
Our results suggest that analogous statements hold for 
radiative corrections given by relativistic Bethe logarithms. 

We have presented results of a calculation of higher-order binding corrections
to the one-loop self-energy for highly excited hydrogenic
atomic levels (see Tab.~\ref{table1}).
Calculational difficulties induced by the 
more complex analytic structure of the wave functions have been a severe
obstacle for evaluations of relativistic Bethe logarithms 
at high $n$, and no prior results are available for 
$A_{60}$ for any state with $n > 4$ (see Ref.~\cite{JeSoMo1997}). 
Intermediate expressions contained up to 200,000 terms; without a computer, this work would have been impractical.
Our calculation is split into a high- and a low-energy part.
We find that the dependence of the low-energy contribution
to $A_{60}$ on the principal quantum number of the 
atomic state under investigation can in many cases be represented
accurately using a three-parameter fit
[see Eq.~(\ref{L1L2L3}) and the data in Tab.~\ref{table2}].
As suggested by the exact formulas for the 
high-energy part given in Eq.~(\ref{Kgen})
and the curves in Figs.~\ref{fig2} and~\ref{fig3},
a fit with less than three parameters cannot be assumed to
lead to a satisfactory representation of~$A_{60}$.
Our final results for~$A_{60}$ are given in Tab.~\ref{table1}.
We establish that the magnitude of $A_{60}$ 
decreases rapidly for Rydberg states with the highest possible
angular momentum for each principal quantum number.
Our calculations 
improve the knowledge of the  self-energy of an electron bound to a nucleus [see Eqs.~(\ref{ESEasF})--(\ref{eq:A60}) and Tab.~\ref{table1}].
They
are motivated by
the dramatically increasing precision of laser
spectrocopy~\cite{BeEtAl1997,ScEtAl1999,ReEtAl2000,NiEtAl2000},
which is rapidly approaching the $1~{\rm Hz}$ level of accuracy.
For the determination of fundamental constants from high-resolution
spectroscopy, frequency measurements of at least two different
transitions have to be performed. Highly excited, slowly 
decaying D states are attractive because they can be excited
out of S states via two-photon resonance~\cite{BeEtAl1997,MoTa2000}.

The authors would like to acknowledge helpful discussions with K.~Pachucki, S.~Jentschura and J.~Sims. U.D.J. acknowledges support from
the Deutscher Akademischer Austauschdienst (DAAD).  E.O.L.
acknowledges support from a Lavoisier fellowship of the French
Ministry of Foreign Affairs, support by NIST, and a grant of computer time at the CINES (Montpellier, France). G.S.~acknowledges
support from BMBF, DFG and from GSI\@.
The Kastler Brossel laboratory is Unit\'e Mixte de Recherche~8552 of the CNRS\@.


\end{document}